\documentclass{emulateapj}
\shortauthors{Wang et al.}

\begin{document}

\title{Near-Infrared Imaging of White Dwarfs with Candidate Debris Disks}

\author{Zhongxiang Wang\altaffilmark{1}, 
Anestis Tziamtzis\altaffilmark{1},
Xuebing Wang\altaffilmark{1,2},
}

\altaffiltext{1}{\footnotesize 
Shanghai Astronomical Observatory, Chinese Academy of Sciences,
80 Nandan Road, Shanghai 200030, China}

\altaffiltext{2}{Graduate University of Chinese Academy of Sciences, 
No. 19A, Yuquan Road, Beijing 100049, China}

\begin{abstract}

We have carried out $JHK_s$ imaging of 12 white dwarf debris disk candidates 
from the WIRED SDSS DR7 catalog, aiming to confirm or rule out disks 
among these sources. On the basis of positional identification and 
the flux density spectra, we find that seven white dwarfs have 
excess infrared emission, but mostly at WISE W1 and W2 bands, four 
are due to nearby red objects consistent with background galaxies or 
very low mass dwarfs, and one exhibits excess emission at $JHK_s$ consistent 
with an unresolved L0 companion at the correct distance. While our 
photometry is not inconsistent with all seven excesses arising from disks, 
the stellar properties are distinct from the known population of debris 
disk white dwarfs, making the possibility questionable. In order to 
further investigate the nature of these infrared sources, warm Spitzer 
imaging is needed, which may help resolve galaxies from the white dwarfs 
and provide more accurate flux measurements.

\end{abstract}


\section{INTRODUCTION}

The discovery of the first circumstellar debris disk around a white dwarf 
(WD; G29-28) was made by \cite{zb87} (see also \citet{jur03, rea+05}). 
Since then, thanks to {\it Spitzer Space Telescope} as well as large 
survey programs, over 20 WDs have been identified to have debris disks 
(see, e.g., \citealt{von+07,fjz09,xj12,gir+12}, and references therein). 
It is believed that such a debris disk is formed from material produced 
by tidal disruption of asteroids within the Roche radius of a 
WD \citep{gra+90,jur03}, since planetary material is known to commonly 
exist around the progenitor stars of WDs and it has been suggested 
that part of the material can survive through late phases of stellar 
evolution (e.g., \citealt{ds02}). 

This picture has been supported by derived properties of several 
gaseous metal disks around isolated WDs \citep{gan+06,gms07,gan+08,mel+12},
detailed infrared studies of the debris-disk WD systems 
(e.g., \citealt{rea+05,von+07,jur+07,rea+09}), and more common detections 
of absorption features of high-Z metals in WD spectra 
(\citealt{zuc+03,kle+11} and references therein). The high-Z spectral 
features are considered as a result of the processes of asteroids 
disruption and subsequent WD accretion of high-Z material probably 
through a disk. The accreted material pollutes the expectedly ``pure" 
hydrogen or helium atmosphere (\citealt{jur08}), since the primordial 
metals within the atmospheres of WDs sink rapidly \citep{paq+86}. 
Therefore, one important application of the observational studies of 
the polluted WDs is that it can provide information about bulk elemental 
compositions of extrasolar planets (see, e.g., \citealt{zuc+07,kle+11,gan+12}).

The large survey of WD debris-disk systems is warranted as the resulting 
increased samples allow detailed studies of disk formation and accretion 
processes around WDs \citep{jur08, raf11}, and that of the disk properties 
and disk-existence frequency indicating the corresponding properties 
of planetary bodies \citep{gir+12,bar+12}. Recently, the Data Release 7 
(DR7) WD catalogue from the Sloan Digital Sky Survey (SDSS), which 
contains nearly 20,000 sources, was released \citep{kle+13}. Using the 
infrared all-sky data from the Wide field Infrared Survey Explorer (WISE), 
\citet{deb+11} conducted a thorough search for infrared counterparts to WDs
in the DR7 catalogue, and found 52 candidate debris disks (we name these 
sources `dxxxx' in this paper) and 69 candidate counterparts with 
indeterminate infrared excess emission (which were defined such because 
both a debris disk and a brown-dwarf companion can explain their excess 
emission; these sources are named as `ixxxx' in this paper). For the first 
and latter types of the counterparts, there are 32 and 54, respectively, 
that did not have detections at $JHK_s$ bands (basically in either 2MASS 
or UKIDSS survey). Since WISE imaging had a FWHM of 
$>$6\arcsec\ \citep{wri+10}, source confusion could cause 
mis-identification of excess emission. In order to identify 
the counterparts among these candidates and if identified, to provide 
more measurements for determining the debris-disk sources, we have 
carried out ground-based imaging of the candidates that did not have 
$JHK_s$ flux measurements. In this paper we report the results from our 
observations.
\begin{figure*}
\begin{center}
\includegraphics[scale=0.90]{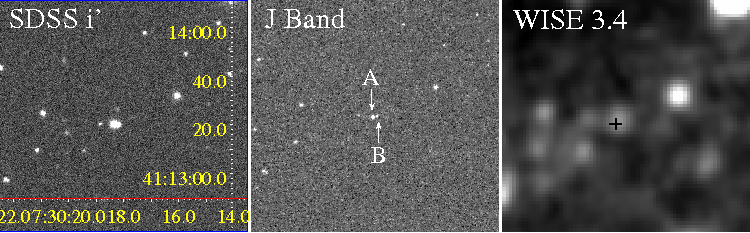}
\includegraphics[scale=0.90]{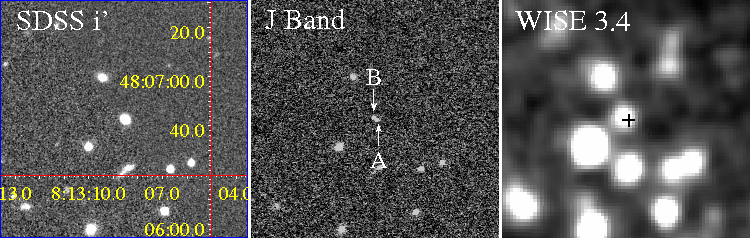}
\includegraphics[scale=0.90]{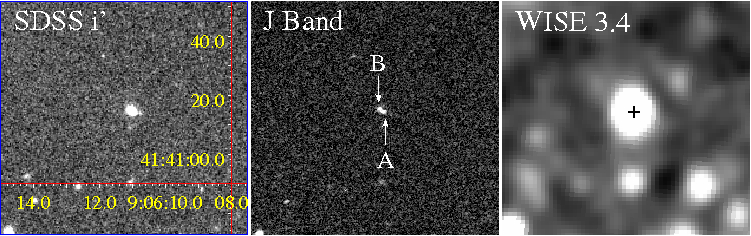}
\includegraphics[scale=0.90]{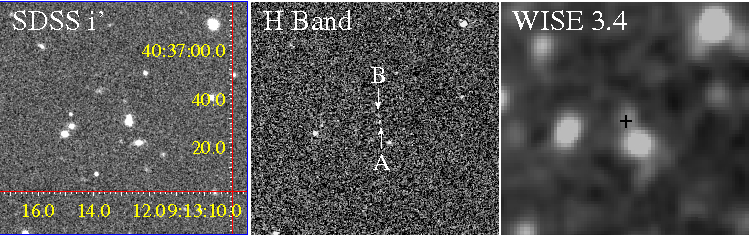}
\caption{Near-infrared images of WDs i0730, d0813, d0906, 
and i0913. These four 
WDs, marked as object $A$ in the middle panels, are resolved to have a nearby 
source (marked as object $B$). The SDSS $i'$ (left panels) and WISE $W1$ band
(right panels) images are shown for comparison. The SDSS positions of the WDs
are marked by plus signs in the WISE images. \label{fig:ds}}
\end{center}
\end{figure*}

\section{OBSERVATIONS AND DATA REDUCTION}    
\label{sec:obs}

\subsection{Ground-based Imaging}

We carried out observations with the 5.1-m Hale telescope at Palomar 
Observatory in the United States. Through the Chinese Telescope Access 
Program, we were awarded two nights in 2012 November and two half nights 
in 2013 February. Unfortunately, only the night of 2012 Nov. 27--28 was 
clear and useful data were taken. The instrument used was the Wide field 
IR Camera (WIRC; \citealt{wil+03}), which has a 2048$\times$2048~pixel$^2$ 
Hawaii-II HgCdTe detector. The pixel scale was 0\farcs249 pixel$^{-1}$ 
and the field of view was 8\farcm7.

For the 86 WDs we chose to observe, we used published results of their 
effective temperature and optical magnitudes to estimate their lower 
flux limits at $JHK_s$ bands (\citealt{deb+11,kle+13}; see also 
Table~\ref{tab:prop}), assuming pure blackbody emission. Based on 
the estimated flux limits, exposure times at the bands for each target 
were estimated accordingly. Due to weather limitations, we observed 
only 12 WDs; the exposure times are given in Table~\ref{tab:obs}. 
During each exposure, the telescope was dithered in a five-point grid 
with offsets of $\sim$ 40\arcsec\ to obtain a measurement of the sky 
background. The observing conditions were mediocre, with the seeing 
having a median value of 0\farcs9 but occasionally dropping to 1\farcs5.

We used IRAF for our data reduction. The images were dark-subtracted and 
flat-field corrected. In addition a sky image was made by filtering 
out stars from each set of the dithered images in one exposure. 
The sky image was subtracted from the set of images, and then 
the sky-subtracted images were shifted and combined into one final image 
of a target field.

To calibrate our target images astrometrically, we used the in-field, 
relatively bright 2MASS stars \citep{2mass}, the numbers of which were 
between eight and fifty, depending on the fields. The resulting nominal 
uncertainties of the calibrated images are in a range of 
0\farcs03--0\farcs11. For most of our images, the uncertainties are 
dominated by the 2MASS systematic uncertainty of $\simeq$ 0\farcs15 
(with respect to the International Celestial Reference System).

We used the IRAF aperture photometry package {\tt apphot} to measure 
the brightnesses of our sources for most of the images. For a few cases 
that a target was resolved to have a nearby source, the PSF fitting 
package {\tt daophot} was used. Flux calibration was conducted by 
comparing to bright 2MASS stars detected in our images.

\subsection{WISE Imaging}

Launched in 2009 December 14, WISE mapped the entire sky 
at 3.4, 4.6, 12, and 22 $\mu$m (called W1, W2, W3, and W4 bands, 
respectively) in 2010 with FWHMs of 6.1\arcsec, 6.4\arcsec, 6.5\arcsec, 
and 12.0\arcsec\  in the four bands, respectively 
(see \citealt{wri+10} for details). The WISE all-sky images and 
source catalogue were released in 2012 March. We downloaded the flux 
measurements of each target in the source catalogue and the WISE image 
data of each target field from the Infrared Processing and Analysis Center.
\citet{deb+11} provided all magnitudes or magnitude upper limits of 
the WISE candidate counterparts to the WD targets, but because they used 
the WISE preliminary catalogues, the values were slightly different from 
those in the all-sky source catalogue. We therefore re-provided 
the magnitudes or magnitude upper limits of our 12 WD targets 
from the all-sky source catalogue in Table~\ref{tab:obs}.

\section{RESULTS}
\label{sec:res}
\subsection{Positional Identification}

In our ground-based images, we detected all 12 targets, but we found that 
four of them were resolved as two sources at/near the WISE source position.
The fields of the four WDs are shown in Figure~\ref{fig:ds}. 
From our astrometry, we determined the counterparts based on the measured 
positions and they are marked as object $A$ in Figure~\ref{fig:ds}. 
The nearby non-counterpart sources, which are 1\arcsec --2\arcsec\ away 
from the counterparts, are marked as object $B$. We also determined  
the positions and $JHK_s$ magnitudes of these nearby sources,
and the values are given next to the counterparts in Table~\ref{tab:obs}. 
For the other eight targets, one single source was clearly detected at 
or near the SDSS position. Among them i0004 was detected by the 2MASS 
survey, but because it had significant proper motion, 
$\Delta\alpha=2\farcs86\pm0\farcs06$ and 
$\Delta\delta=-0\farcs22\pm0\farcs04$ (our Palomar measurement with 
respect to that of 2MASS, which was made on 1998 Sept. 17), it was not 
reported to have the 2MASS detections (the positional criterion for 
candidate counterpart identification was 2\farcs0 in \citealt{deb+11}).

\subsection{Flux Density Spectra}
\begin{figure*}
\centering
\begin{tabular}{c c}
\includegraphics[scale=0.48]{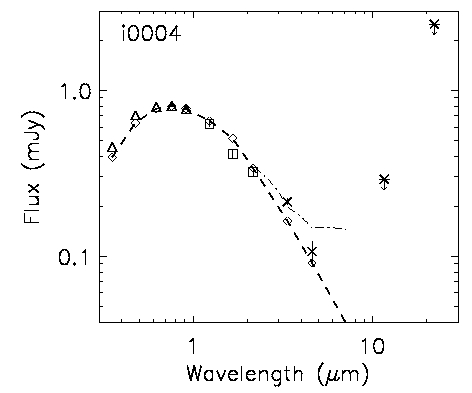} 
& \includegraphics[scale=0.48]{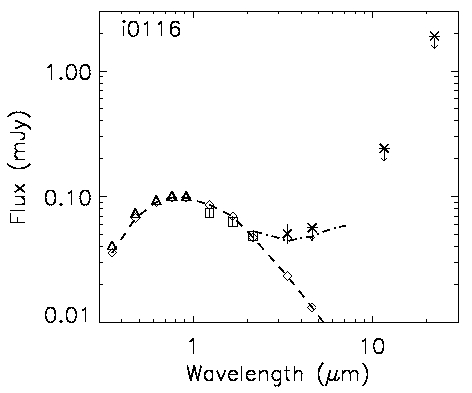}\\
 \includegraphics[scale=0.48]{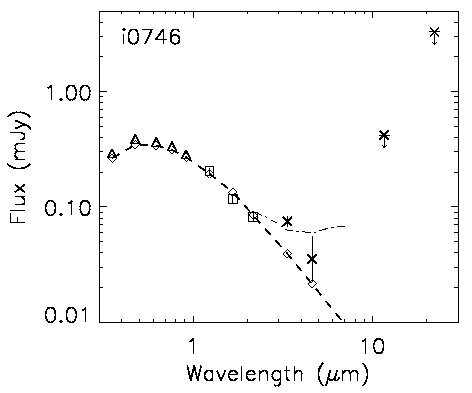} &
 \includegraphics[scale=0.48]{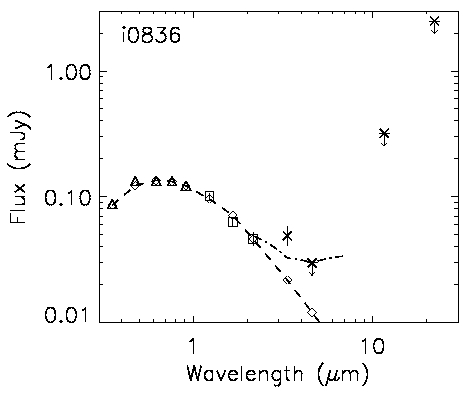}\\
 \includegraphics[scale=0.48]{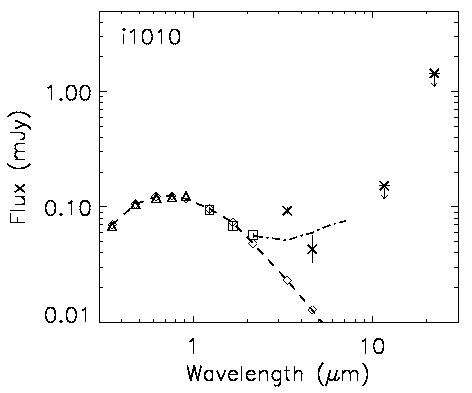} &
\includegraphics[scale=0.48]{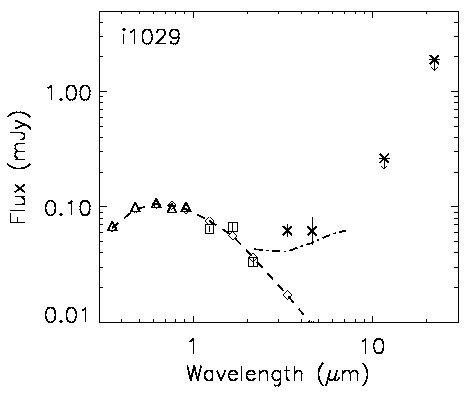} \\ 
 \includegraphics[scale=0.48]{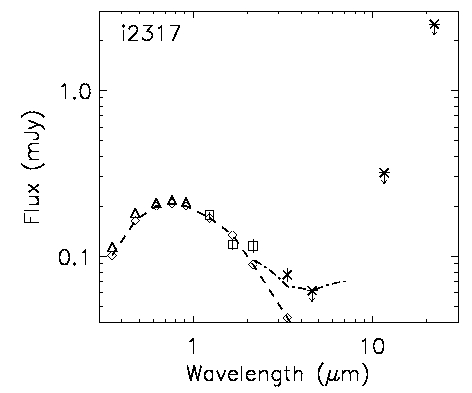} & \\
\end{tabular}
\caption{Flux density spectra of 7 WDs that possibly have a debris disk. The
SDSS optical, our Palomar $JHK_s$, and WISE $W1$/$W2$ fluxes 
are displayed as triangles, squares, and crosses, respectively. The WISE flux
upper limits are also shown. The model fluxes of each WD at the bands
are indicated by diamonds and connected by the dashed curve, and the best-fit
debris disk model spectrum is plotted as the dash-dotted curve. \label{fig:dd}}
\end{figure*}

\begin{figure*}
\centering
\begin{tabular}{c c}
\includegraphics[scale=0.56]{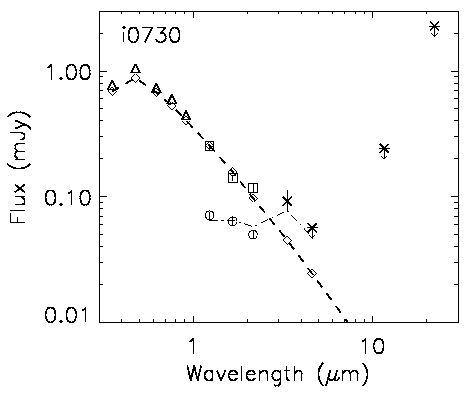} & 
\includegraphics[scale=0.56]{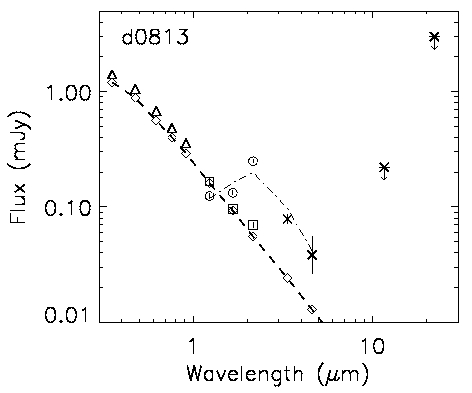} \\
\includegraphics[scale=0.56]{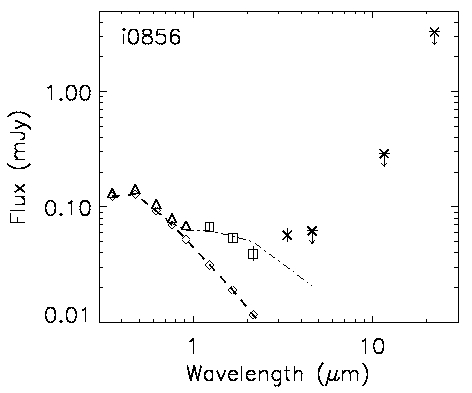} & 
 \includegraphics[scale=0.56]{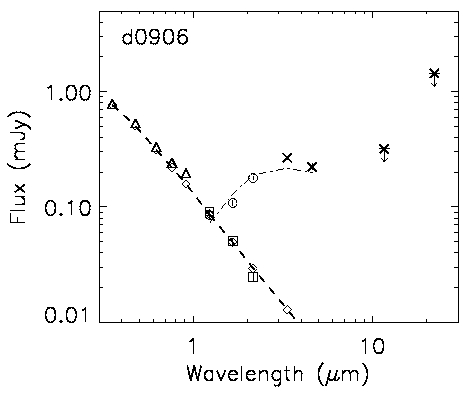}\\
\includegraphics[scale=0.56]{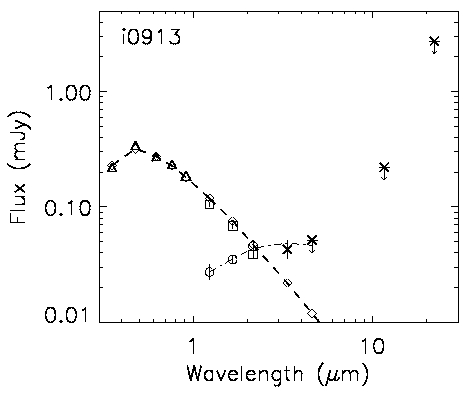} &  \\
\end{tabular}
\caption{Flux density spectra of 5 WDs that possibly have either a VLM dwarf 
nearby in the field or a possible dwarf companion (only for i0856).
Symbols are the same as in Figure~\ref{fig:dd}, except with the 
dash-dotted curve indicating the best-fit brown dwarf model spectrum and 
the circular data points the fluxes of object $B$ in Figure~\ref{fig:ds}.
\label{fig:bd}}
\end{figure*}

Combining SDSS $u'g'r'i'z'$ flux measurements \citep{kle+13} and that 
from the WISE all-sky source catalogue with our $JHK_s$ measurements, 
we constructed the flux density spectra for the 12 WD targets. The 
spectra are shown in Figures~\ref{fig:dd} and \ref{fig:bd}. For 
the four WDs with a nearby source, the nearby sources are included 
correspondingly in the figure (displayed as circular data points). 
We compared our observational spectra with WD model spectra in 
the infrared bands (kindly provided by P. Bergeron), whose properties 
were determined by \citet{kle+13} (see also \citealt{deb+11}), and 
found that no significant excess emission at $JHK_s$ bands was detected 
for most of the WDs except i0856. For i0730, d0813, d0906 and i0913, 
the emission detected by WISE more likely came from their nearby source 
(see Section~\ref{sssec:bd} below), and our observations excluded 
them as the WDs with excess infrared emission.

\section{Discussion and Summary}
\label{sec:disc}

We observed 12 WDs that were identified to have excess emission from 
the WISE survey by \citet{deb+11} but did not have previous $JHK_s$ 
measurements. Given the excess emission, they have been suggested 
to either have a debris disk or a brown dwarf companion \citep{deb+11}. 
From our observations, we found that seven of them did not have 
significant excess emission at $JHK_s$ bands, while i0856 had strong 
excess emission, consistent with fluxes measured at the SDSS $r'i'z'$ 
bands. In addition four WDs were resolved to have a nearby source. 
Below including our results, we first discuss the possible origins 
for the excess emission from the WDs and for the resolved nearby 
sources, and then provide a summary for the discussion.

\subsection{Candidate debris disk sources?}

For the seven WDs without significant $JHK_s$ excess emission, 
they are not likely to have a very low-mass (VLM) star or a brown dwarf 
companion. For example, the $K_s$ magnitudes and their uncertainties 
are 15--18 and $\sim$0.1, respectively. The uncertainties only allow 
the possible existence of 10\% excess emission or 2.5~mag fainter sources. 
Adding 2.5~mag to $K_s$ and comparing it to $W1$ magnitudes, which are 
slightly lower than $K_s$ values (see Table~\ref{tab:obs}), such 
infrared sources would have a $K_s-W1$ color of $>$2.5 mag. The color 
is too red for VLM dwarfs (see \citealt{kir+11} for the colors of 
the known VLM dwarfs). 

With constraints from our $JHK_s$ measurements, we fit these sources 
with the debris disk model given by \citet{jur03} to study if 
the excess emission could arise from a debris disk. In the model, 
the disk temperature follows $T(r)\propto T_{\rm WD} r^{-3/4}$, 
where $T_{\rm WD}$ is the effective temperature of a WD and $r$ 
the disk radius \citep{jur03}. We adopted the model parameters used 
in \citet{deb+11}, assuming a temperature of $T=1200$~K at the inner edge 
of the debris disk and an outer disk radius of 80$r_{\rm WD}$, 
where $r_{\rm WD}$ is the radius of a WD. The distance, effective 
temperature, and extinction of each WD target were fixed at the values 
given in Table~\ref{tab:prop} \citep{deb+11,kle+13}. The free parameter 
was the inclination angle $i$ of the disk. We fit $K_s$, $W1$, and $W2$ 
(or only $K_s$ and $W1$ when there was no $W2$ detection) fluxes, 
where $K_s$ was included to serve as the additional constraint 
(for i2317, which had $K_s$ excess emission, $H$ band flux was included too). 
We also required that the model flux at $W2$ band must be lower than 
the WISE flux upper limit for the sources not detected at $W2$. We found 
that the excess emission from most of the sources is generally  
consistent with arising from a debris disk, and the resulting best-fit 
$\cos i$ and $\chi^2$ values are summarized in Table~\ref{tab:dd}. 
The best-fit model fluxes for each source are shown in Figure~\ref{fig:dd}.

For i0836 and i2317, the requirement of the model for the $W2$ flux lower 
than the WISE upper limit provided a constraint in the fitting. 
If this requirement is not considered, $\chi^2$ values would be much 
smaller. For i1010 and i1029, the $\chi^2$ values are quite large 
for $\cos i=1$. This is because their $W1$ fluxes are significantly 
higher than their $K_s$ fluxes, and in order for the model to match 
the $W1$ fluxes, the model fluxes were increased, thus inducing 
large $\chi^2$ values at $K_s$ band. We note that since the WISE 
magnitudes of the WDs are in a range of 15--17 and WISE photometry 
of such faint sources is known to have as large as $\sim$0.4 mag systematic
uncertainty\footnote{see http://wise2.ipac.caltech.edu/docs/release/allsky/expsup/sec6\_3c.html}, 
the poor fitting can be caused by the large uncertainties on WISE 
photometry that are not included in the catalogue data. For the same 
reason, we did not further search for better fitting by varying 
the model parameters.

However for the WDs that show clear excess emission and are believed 
to have a debris disk, we know that: 1) they nearly all have effective 
temperatures in a range of 9,500--24,000 K (except G~166$-$58 for 
its $T_{\rm WD}=7400$ K; \citealt{fzb08}), 2) they nearly all are known 
to be metal-rich from  optical spectroscopy \citep{xj12}, 
and 3) nearly half of them show significant emission excesses at infrared 
$K$-band relative to their WD model spectra. These properties make 
the identification of the seven WDs as candidates with debris disks 
highly questionable. The seven WDs generally do not 
fit in any of them (Tables~\ref{tab:prop} \& \ref{tab:dd}; \citealt{deb+11}).

\subsection{Candidate brown dwarf sources?}
\label{sssec:bd}

Since WISE imaging had relatively low spatial resolution, it can not be 
determined solely based on the positions whether or not the WISE sources 
are the counterparts to the nearby sources of i0730, d0813, d0906, 
and i0913 (object $B$ in Figure~\ref{fig:ds}). Combining our $JHK_s$ 
measurements with the WISE fluxes, as shown in Figure~\ref{fig:bd}, 
the overall broad-band spectra suggest that the WISE sources are 
the counterparts to the four nearby sources, or at least emission from 
the nearby sources dominated over that from the four WDs 
(otherwise these nearby sources would have to have an unlikely, 
large flux decrease from $K_s$ to $W1$). Since the sources are red 
and three of them (nearby to ixxxx sources) were classified 
by \citet{deb+11} to be possible candidate brown dwarfs, we fit 
their broad-band spectra with that of VLM stars and brown dwarfs. 
Following \citet{deb+11}, we used the empirical spectra for M, L, and 
T dwarfs \citep{haw+02,kir+11}. We first fixed the distances at the values
of the WDs, and found that the resulting $\chi^2$ were large 
(the values and the best-fit spectral types are given in column four 
and three, respectively, of Table~\ref{tab:bd}). Setting the distance 
as a free parameter, the best-fit $\chi^2$ can be significantly reduced 
(see the values at column seven of Table~\ref{tab:bd}). Therefore, if 
the sources are M or L dwarfs, as identified from our fitting, they are 
most likely not associated with the WDs. The best-fit spectra of 
the four sources are shown in Figure~\ref{fig:bd}.

In addition for i0856, since it had significant excess emission starting 
from the optical $z'$ to $JHK_s$ bands comparing to the WD model spectrum, 
the debris-disk model we used, which assumed a low-temperature disk, 
could not provide a reasonably well fit. We thus tested with the VLM dwarf 
models, and found that an L0 dwarf at the WD's distance of 428 pc can 
generally describe the excess emission except at the $W1$ band 
(Figure~\ref{fig:bd}). Here again given the large uncertainties 
on the WISE measurements of faint sources, we considered the fitting 
was acceptable although the reduced $\chi^2$ is $\simeq 16$ 
(Table~\ref{tab:bd}).

\subsection{Background galaxies?}

The WD targets are located away from the Galactic plane having clean 
source fields according to the SDSS optical and our $JHK_s$ images. 
However, it has been shown from \textit{Spitzer} imaging that 
at $m > 14$ mag in the wavelength range of 3--10~$\mu$m, galaxies 
dominate in such regions \citep{faz+04}. For example as our WD targets 
have $W1$ magnitudes in a range of 15--17 mag, at its middle value of 
16~mag, the \textit{Spitzer} galaxy count at 3.6 $\mu$m was 
$\simeq$2300 mag$^{-1}$~deg$^{-2}$ \citep{faz+04} 
or 1.8$\times 10^{-4}$ mag$^{-1}$~arcsec$^{-2}$. 
Considering the 2\arcsec\ radius circular region, which was used 
by \citet{deb+11} for searching for WD counterparts, there will be a 
chance of 0.23\% to randomly find at least one 16 mag galaxy in such a 
region. The percentage is low but there were nearly 18,000 times 
searches (for 17,955 unique and valid targets; \citealt{deb+11}), 
which would result in 40 randomly-detected galaxies. Approximately 
300 WDs per mag were detected at $W1=16$ mag, but excluding 
23\% naked WDs (detection of WD photosphere only) and 
67\% candidate WD plus M dwarf binaries \citep{deb+11}, the latter we 
consider rather certain due to their brightnesses and colors, only 
30 per mag would be either debris-disk or brown-dwarf companion systems 
among the candidates. The numbers thus suggest that the detected excess 
emission is likely caused by unresolved background galaxies due 
to WISE's relatively low spatial resolution and those nearby sources 
are also likely galaxies.

\subsection{Summary}

Among the 12 WD targets identified with excess emission from 
the WISE data, our observations and analysis show that seven are 
consistent with having a debris disk, but their properties are not 
in the likely range for the detectable disks on the basis of 
the currently known debris-disk WDs. Among the other five WD targets, 
four are found that their excess emission is caused by the existence 
of a nearby red source and the remaining one, i0856, shows significant 
excess emission at $JHK_s$ bands. Our analysis suggests that the nearby 
sources are possibly unassociated VLM stars or brown dwarfs while 
excess emission from i0856 is suggestive of an L0 dwarf. However, we 
also realize that the excess emission (and the nearby sources) might 
well be caused by background galaxies, which are known to be the dominant, 
relatively faint sources at wavelengths between 3--10 $\mu$m. 
Therefore, in order to investigate the true nature of the observed 
excess emission or nearby sources, imaging with \textit{Spitzer} is 
needed. The \textit{Spitzer} observations will possibly resolve 
background galaxies from the WD targets and provide accurate flux 
measurements at the infrared wavelengths of $W1$ and $W2$ bands, 
both helping identify the debris disks and VLM dwarfs.

\acknowledgements

This research uses data obtained through the Telescope Access Program 
(TAP), which is funded by the National Astronomical Observatories, 
Chinese Academy of Sciences, and the Special Fund for Astronomy from 
the Ministry of Finance. This publication makes use of data products 
from the Two Micron All Sky Survey, which is a joint project of 
the University of Massachusetts and  the Infrared Processing and 
Analysis Center/California Institute of Technology, funded by the National 
Aeronautics and Space Administration and the National Science Foundation. 
The publication also makes use of data products from the Wide-field 
Infrared Survey Explorer, which is a joint project of the University 
of California, Los Angeles, and the Jet Propulsion Laboratory/California 
Institute of Technology, funded by NASA.

We gratefully thank anonymous referee for very constructive suggestions, 
and P. Bergeron for providing us the WD model spectrum data. This research 
was supported by National Basic Research Program of China 
(973 Project 2009CB824800), and National Natural Science Foundation of 
China (11073042, 11373055). Z.W. is a Research Fellow of the 
One-Hundred-Talents project of Chinese Academy of Sciences. 
A.T. acknowledges support from Chinese Academy of Sciences visiting 
Fellowship for Researchers from Developing Countries.

{\it Facilities:} \facility{Hale (WIRC)} 

\bibliographystyle{apj}

\begin{deluxetable}{l c c c c c c c c c}
\tablecolumns{10}
\tablecaption{Properties of 12 WD targets \label{tab:prop}}
\tablewidth{0pt}
\tablehead{
\colhead{Name} & \colhead{$u'$} & \colhead{$g'$} & \colhead{$r'$} &
\colhead{$i'$} & \colhead{$z'$} & \colhead{$T$ (K)} & \colhead{$\log(g)$\tablenotemark{a}} & \colhead{$A_{g'}$} & \colhead{$d$ (pc)}}
\startdata
i000410.42$-$034008.6 & 17.46 & 16.93 & 16.76 & 16.72 & 16.74 & 6887 & 7.71 & 0.15 & 51 \\
i011616.95$-$094347.9 & 20.09 & 19.39 & 19.08 & 18.97 & 18.95 & 6309 & 8.06 & 0.16 & 104\\
i073018.36$+$411320.4 & 17.03 & 16.61 & 16.92 & 17.10 & 17.38 & 15126 & 7.83 & 0.26 & 133 \\
i074631.42$+$173448.2 & 17.93 & 17.57 & 17.59 & 17.66 & 17.82 & 9282 & 8.59 & 0.14 & 66 \\
d081308.52$+$480642.3 & 16.30 & 16.54 & 16.97 & 17.30 & 17.60 & 32727 & 7.86 & 0.20 & 279 \\
i083633.00$+$374259.4 & 19.21 & 18.71 & 18.67 & 18.66 & 18.73 & 7798 & 8.11 & 0.12 & 116 \\
i085650.58$+$275118.0 & 18.76 & 18.63 & 18.92 & 19.21 & 19.35 & 19333 & 7.86 & 0.12 & 428 \\
d090611.00$+$414114.3 & 16.73 & 17.14 & 17.64 & 17.97 & 18.19 & 47637 & 7.91 & 0.05 & 469 \\
i091312.74$+$403628.8 & 18.14 & 17.64 & 17.86 & 18.02 & 18.25 & 11726 & 8.02 & 0.07 & 153 \\
i101007.89$+$615515.7 & 19.38 & 18.89 & 18.73 & 18.71 & 18.68 & 7252 & 8.31 & 0.04 & 94 \\
i102915.97$+$300251.6 & 19.42 & 18.98 & 18.88 & 18.97 & 18.93 & 7755 & 7.86 & 0.08 & 153 \\
i231725.29$-$084032.9 & 18.94 & 18.38 & 18.19 & 18.13 & 18.14 & 6862 & 7.34 & 0.13 & 124 \\
\enddata
\tablenotetext{a}{$\log(g)$ is in units of cm s$^{-2}$.}
\end{deluxetable}

\begin{deluxetable}{l c c c c c c c c c c }
\tabletypesize{\small}
\tablecolumns{11}
\tablecaption{Near-infrared and WISE measurements of 12 WD targets\tablenotemark{a,b}\label{tab:obs}}
\tablewidth{0pt}
\tablehead{\colhead{Name} & \multicolumn{2}{c}{$J$ Filter} & 
\multicolumn{2}{c}{$H$ Filter} & \multicolumn{2}{c}{$K_s$ Filter} &
\colhead{W1} & \colhead{W2} & \colhead{W3} & \colhead{W4} \\
\colhead{} & \colhead{$t_{\rm exp}$} & \colhead{$J$} & 
\colhead{$t_{\rm exp}$} & \colhead{$H$} & 
\colhead{$t_{\rm exp}$} & \colhead{$K_s$} &
\colhead{} & \colhead{} & \colhead{} & \colhead{} } 
\startdata
i000410.42$-$034008.6 & 0.27 & 16.05$\pm0.04$ & 0.27 & 16.01$\pm$0.07 & 0.27 & 15.80$\pm$0.07 & 15.40$\pm$0.05 & 15.51$\pm$0.2 & 12.5 & 8.8 \\
i011616.95$-$094347.9 & 5.0 & 18.38$\pm$0.09 & 5.0 & 18.05$\pm$0.10 & 5.0 & 17.86$\pm$0.11 & 17.0$\pm$0.2 & 16.2 & 12.7 & 9.1 \\
i073018.36$+$411320.4 & 0.83 & 17.07$\pm$0.04 & 1.7 & 17.20$\pm$0.06 & 2.5 & 16.91$\pm$0.08 & 16.3$\pm$0.2 & 16.2 & 12.7 & 8.9 \\
073018.20$+$411320.4\tablenotemark{c} &  & 18.44$\pm$0.06 &  & 18.06$\pm$0.08 &  & 17.84$\pm$0.07 &                &                &      &     \\
i074631.42$+$173448.2 & 1.3 & 17.26$\pm$0.05 & 1.7 & 17.37$\pm$0.08 & 2.5 & 17.29$\pm$0.09 & 16.5$\pm$0.1 & 16.7$\pm$0.5 & 12.1 & 8.5 \\
d081308.52$+$480642.3 & 1.3 & 17.50$\pm$0.04 & 2.5 & 17.60$\pm$0.06 & 5.0 & 17.47$\pm$0.06 & 16.5$\pm$0.1 & 16.6$\pm$0.4 & 12.8 & 8.6 \\
081308.61$+$480643.4\tablenotemark{c} &  & 17.82$\pm$0.06  &  & 17.25$\pm$0.05 & & 16.09$\pm$0.04 &                &                &      &     \\
i083633.00$+$374259.4 & 2.5 & 18.03$\pm$0.06 & 5.0 & 18.05$\pm$0.08 & 10 & 17.92$\pm$0.13 & 17.0$\pm$0.2 & 17.0 & 12.4 & 8.8 \\
i085650.58$+$275118.0 & 7.2 & 18.27$\pm$0.09 & 16.2 & 18.30$\pm$0.09 & 27 & 18.41$\pm$0.16 & 16.8$\pm$0.2 & 16.2 & 12.5 & 8.5 \\
d090611.00$+$414114.3 & 2.5 & 18.12$\pm$0.05 & 5.0 & 18.28$\pm$0.07 & 7.5 & 18.59$\pm$0.09 & 15.15$\pm$0.04 & 14.71$\pm$0.07 & 12.4 & 9.4\\
090611.09$+$414115.1\tablenotemark{c}&  & 18.20$\pm$0.05 &  & 17.45$\pm$0.05 &  & 16.44$\pm$0.04 &                &                &      &     \\
i091312.74$+$403628.8 & 2.5 & 17.97$\pm$0.05 & 2.5 & 17.95$\pm$0.06 & 5.0 & 18.10$\pm$0.09 & 17.1$\pm$0.2 & 16.3 & 12.8 & 8.7\\
091312.73$+$403631.3\tablenotemark{c} &  & 19.44$\pm$0.16 &  & 18.68$\pm$0.07 &  & 17.92$\pm$0.08 &                &                &       &    \\
i101007.89$+$615515.7 & 2.5 & 18.09$\pm$0.07 & 5.0 & 17.95$\pm$0.08 & 5.0 & 17.68$\pm$0.05 & 16.34$\pm$0.07 & 16.5$\pm$0.3 & 13.2 & 9.4 \\
i102915.97$+$300251.6 & 5.0 & 18.51$\pm$0.06 & 5.0 & 17.97$\pm$0.10 & 7.5 & 18.27$\pm$0.10 & 16.7$\pm$0.1 & 16.1$\pm$0.3 & 12.6 & 9.1 \\
i231725.29$-$084032.9 & 3.3 & 17.42$\pm$0.05 & 3.3 & 17.37$\pm$0.08 & 2.5 & 16.92$\pm$0.10 & 16.5$\pm$0.1 & 16.1 & 12.4 & 8.8 \\
\enddata
\tablenotetext{a}{Exposure time $t_{\rm exp}$ at each band is in units of minute.}
\tablenotetext{b}{For WISE measurements, magnitudes without uncertainties are upper limits.}
\tablenotetext{c}{Nearby stars that are marked as object $B$ in Figure~\ref{fig:ds}, with their names giving the positions measured by our near-infrared imaging.}
\end{deluxetable}

\begin{deluxetable}{lcc}
\tablecolumns{4}
\tablecaption{Results from debris-disk fitting\label{tab:dd}}
\tablewidth{0pt}
\tablehead{\colhead{Name} & \colhead{$\cos i$} & \colhead{$\chi^2$/DoF} }
\startdata
i0004 & 0.18 &  7.5/2 \\
i0116 & 0.93 &  1.1/1 \\
i0746 & 0.32  &  1.5/2  \\
i0836 & 0.30 &  7.4/1 \\
i1010 & 1.0  &  46/2  \\
i1029 & 1.0 &   20/2  \\
i2317 & 0.46 &  11/2 \\
\enddata
\end{deluxetable}

\begin{deluxetable}{lcccccc}
\tablecolumns{7}
\tablecaption{Results from VLM dwarf fitting\label{tab:bd}}
\tablewidth{0pt}
\tablehead{\colhead{Name} & \colhead{$d$\tablenotemark{a} (pc)} & \colhead{Spectral Type} & \colhead{$\chi^2/$DoF} & \colhead{$d$\tablenotemark{b} (pc)} & \colhead{Spectral Type} & \colhead{$\chi^2/$DoF} }
\startdata
i0730B & 133 & L4  & 170/4 & 457 & M6 & 9.1/3 \\
d0813B & 279 & M6 & 176/4 & 166 &  L0 & 59/3 \\
i0856  & 428 & L0 & 97/6  & \nodata & \nodata & \nodata \\
d0906B & 469 & M5 & 706/4  & 110 &  L5 & 52/3 \\
i0913B & 153 & L6 & 28/3  & 87 &   L9 & 0.85/2 \\
\enddata
\tablenotetext{a}{Distance fixed at that of the corresponding WD.}
\tablenotetext{b}{Obtained distance when it is set as a free parameter.}
\end{deluxetable}

\end{document}